\documentclass[onecolumn,showpacs,showkeys,preprintnumbers,amsmath,amssymb]{revtex4}
\usepackage{amssymb}
\usepackage{graphicx}
\usepackage{subfigure}
\usepackage[dvipdfm]{hyperref}

\begin{document}

\title{Eliminating interactions between non-neighboring qubits in the
preparation of cluster states in quantum molecules}
\author{Guo-Ping Guo}
\email{gpguo@ustc.edu.cn}
\author{Xiao-Jie Hao}
\author{Tao Tu}
\email{tutao@ustc.edu.cn}
\author{Zhi-Cheng Zhu}
\author{Guang-Can Guo}
\affiliation{Key Laboratory of Quantum Information, University of
Science and Technology of China, Chinese Academy of Sciences, Hefei
230026, People's Republic of China}
\date{\today}

\begin{abstract}
We propose a scheme to eliminate the effect of non-nearest-neighbor
qubits in preparing cluster state with double-dot molecules. As the
interaction Hamiltonians between qubits are Ising-model and mutually
commute, we can get positive and negative effective interactions
between qubits to cancel the effect of non-nearest-neighbor qubits
by properly changing the electron charge states of each quantum dot
molecule. The total time spent on the present multi-step cluster
state preparation scheme is only doubled for one-dimensional qubit
chain and tripled for two-dimensional qubit array comparing with the
time spent on previous protocol leaving out the non-nearest-neighbor
interactions.
\end{abstract}

\maketitle

Semiconductor quantum dot (QD) is believed to be one of the most
promising systems for quantum information processing
\cite{Nielsen,Loss}. Recently, quantum molecule, formed by coupled
double quantum dots, has attracted many interests. A lot of
experimental \cite{Petta, Koppens Nature, Koppens Science, Marcus arxiv}
 and theoretical \cite{Taylor,J.M. Taylor
et al.,optical} work has be done on the two-electron states of the
double-dot molecule. It is argued that by encoding in singlet and
triplet states of quantum molecule, qubits can be protected from
low-frequency noise and the dominant source of decoherence from
hyperfine interactions can be suppressed \cite{Wu,Taylor prl,J.R.
Petta arxiv,Johnson,de Sousa,Petta}. Universal quantum gates and a
fault-tolerant architecture for quantum computation have been
proposed for these qubits encoding in the two-electron states of
quantum molecule \cite{Hanson,J.M. Taylor et al.}. In the reference
\cite{Guo}, we proposed a one-step scheme to prepare large cluster
states with QD molecules, which can act as a general source for
one-way quantum computation \cite{cluster}. As in most of the
solid-state quantum computation protocols, only interactions between
neighboring qubits are considered in this scheme. However, all these
protocols directly explore the long range Coulomb interaction of the
electrons between different molecules \cite{Hanson,J.M. Taylor et
al.,Guo}. There are unwanted small interactions between
non-nearest-neighbor molecules. Generally, this kind of unwanted
interactions between non-nearest-neighbor qubits will act as an
additional decoherence source and affect the proposed quantum
processing fidelity, for example, the fidelity of prepared states or
quantum gates. It is well known that there is a decoupling and
refocusing technique in standard nuclear-magnetic-resonance field
\cite{Slichter,NMR}. Many works explores this idea to simulate
quantum dynamics \cite{Simulation}, construct universal quantum
computation with some Hamiltonian forms \cite{Jennifer}. Here we
focused on a particular physical system of double-dot quantum
molecules, and propose an efficient scheme to eliminate the effect
of the interactions between non-nearest-neighbor qubits in the
preparation of cluster state with QD molecules. By properly changing
the electron charge states in each QD molecule, the effective
interactions between qubits can be switched as positive, zero and
negative. As the interaction Hamiltonians between qubits are all
Ising-model and mutually commute, the effect of non-nearest-neighbor
qubits interactions can be efficiently canceled through a multi-step
preparation process. In addition, the total time for the present
multi-step cluster state preparation scheme is only several times
comparing with the time of previous protocol neglecting the
non-nearest-neighbor interactions.

\begin{figure}[tb]
\includegraphics[width=0.85\columnwidth]{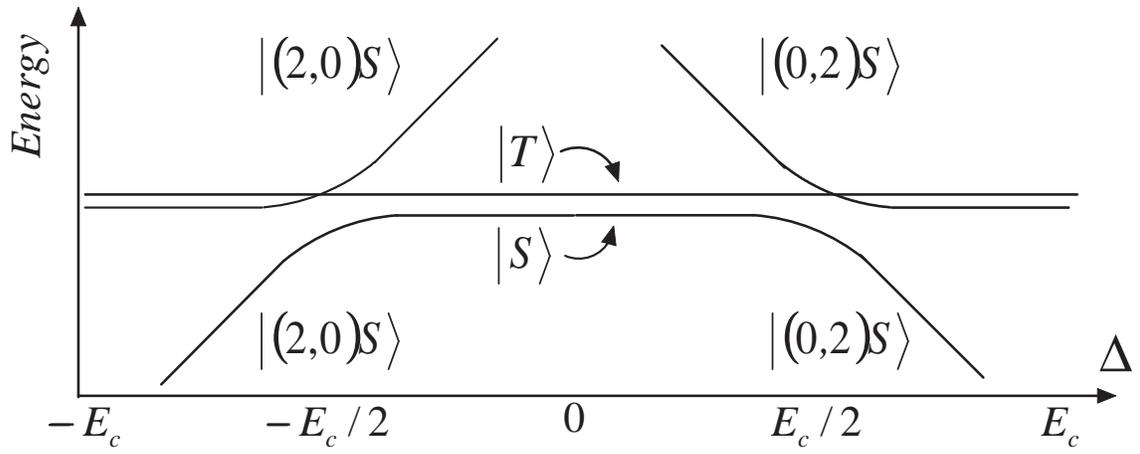}
\caption{Energy level structure of the quantum molecule system as a
function of potential offset $\Delta $ between the three possible
charge states.} \label{fig:1}
\end{figure}

For each double-dot quantum molecule, the notation $(n_{u},n_{l})$
can be used to indicate the number of electrons in the upper and
lower QDs. Define a bias parameter $\Delta $ to represent the
potential offset between the three possible charge states of each
molecule $(0,2)$, $(1,1)$ and $(2,0)$. We can adiabatically sweep
$\Delta $ of each molecule by tuning gate-bias voltages of each
molecule or applying an external electrical field. The charge states
$(0,2)$, $(1,1)$ and $(2,0)$ respectively correspond to the case of
$\Delta $ $=E_{c},0$ and $-E_{c}$ as shown in Fig.\ref{fig:1}
\cite{Petta,Taylor}.

For the charge state $(1,1)$, there are four spin states:
$\left\vert S\right\rangle =(\left\vert \uparrow \downarrow
\right\rangle -\left\vert \downarrow \uparrow \right\rangle
)/\sqrt{2}$, $\left\vert T\right\rangle =(\left\vert \uparrow
\downarrow \right\rangle +\left\vert \downarrow \uparrow
\right\rangle )/\sqrt{2}$, $\left\vert T_{+}\right\rangle
=\left\vert \uparrow \uparrow \right\rangle $, and $\left\vert
T_{-}\right\rangle =\left\vert \downarrow \downarrow \right\rangle
$. The two triplet states $\left\vert T_{\pm }\right\rangle $ can be
largely separated from the singlet state $\left\vert S\right\rangle
$ and triplet state $\left\vert T\right\rangle $ by Zeeman
splitting, which enables us to exclude them in the following
\cite{J.M. Taylor et al.}. The qubit can be encoded in the
near-degenerate $(1,1)$ charge states $\left\vert T\right\rangle
=\left\vert 0\right\rangle $ and $\left\vert S\right\rangle
=\left\vert 1\right\rangle $, which are separated due to weak
tunneling between the two dots of each molecule. As a result of
Pauli blockade, only the singlet state $\left\vert S\right\rangle $
can be tuned into charge state $(0,2)$ or $(2,0)$ by sweeping
$\Delta $. The triplet state $\left\vert T\right\rangle $ will be
kept in the charge state $(1,1)$ in the processing of sweeping
$\Delta $.

\begin{figure}[tb]
\subfigure []
{\label{fig:2:a}\includegraphics[width=0.35\columnwidth]{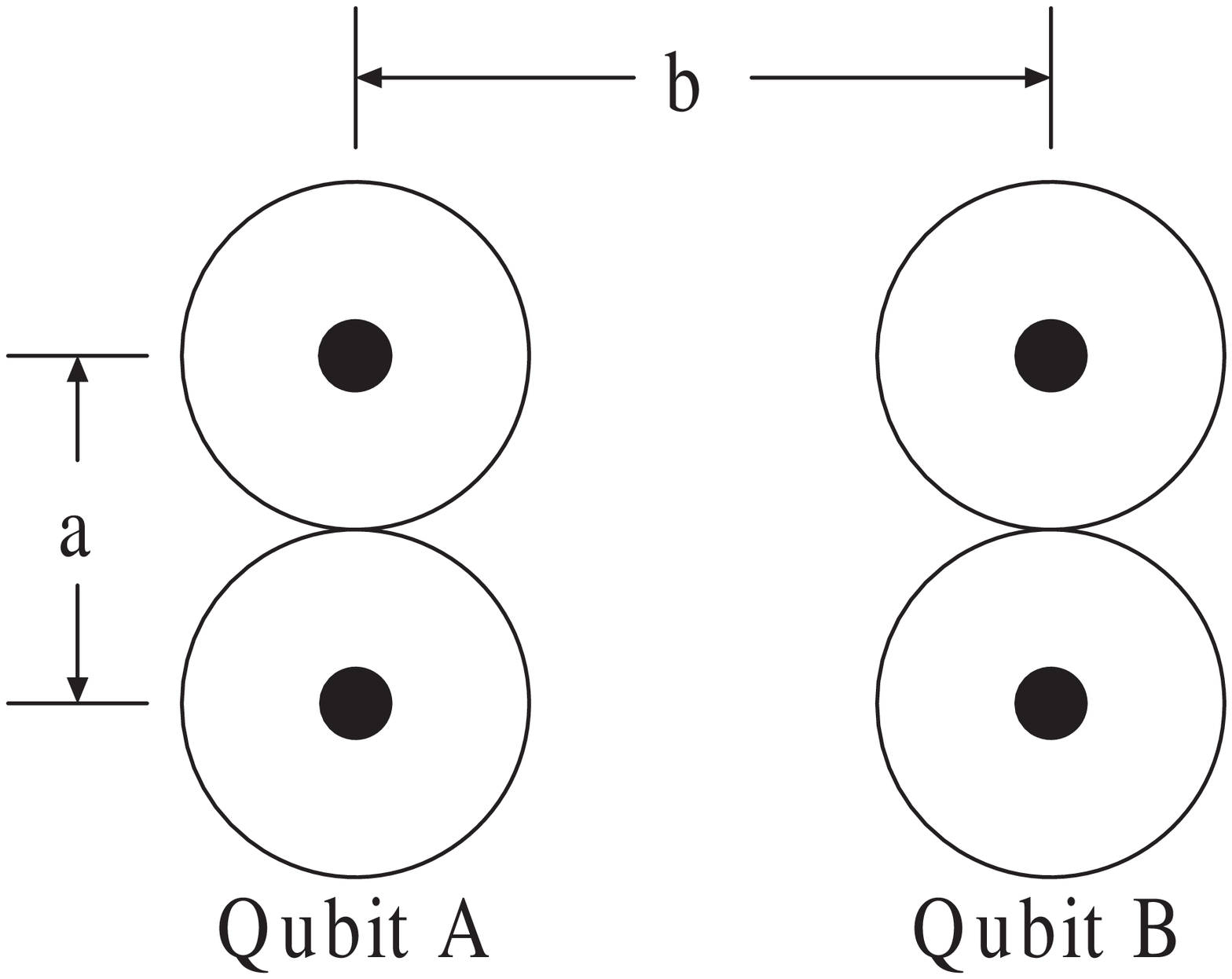}} %
\subfigure []
{\label{fig:2:b}\includegraphics[width=0.35\columnwidth]{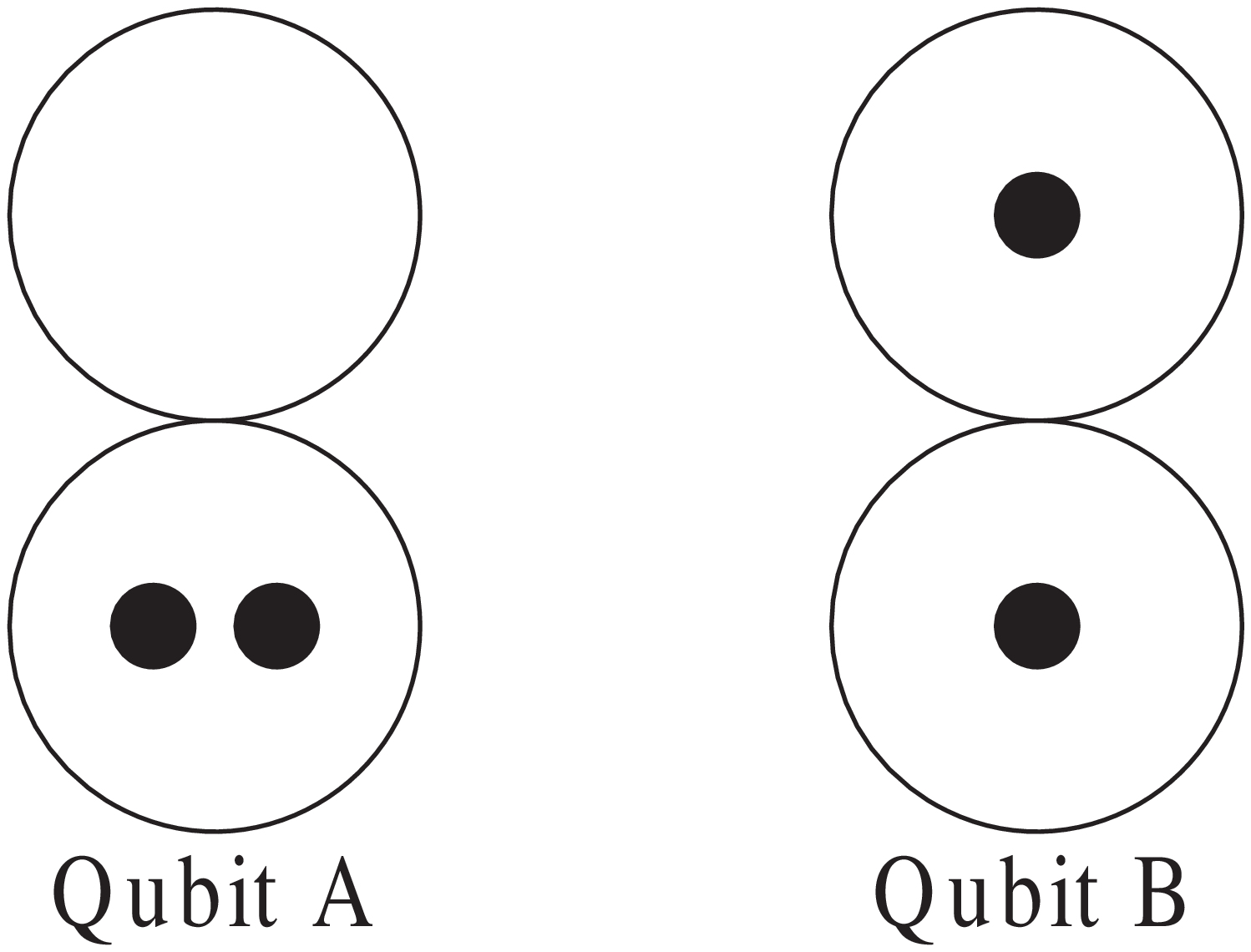}} %
\subfigure []
{\label{fig:2:c}\includegraphics[width=0.35\columnwidth]{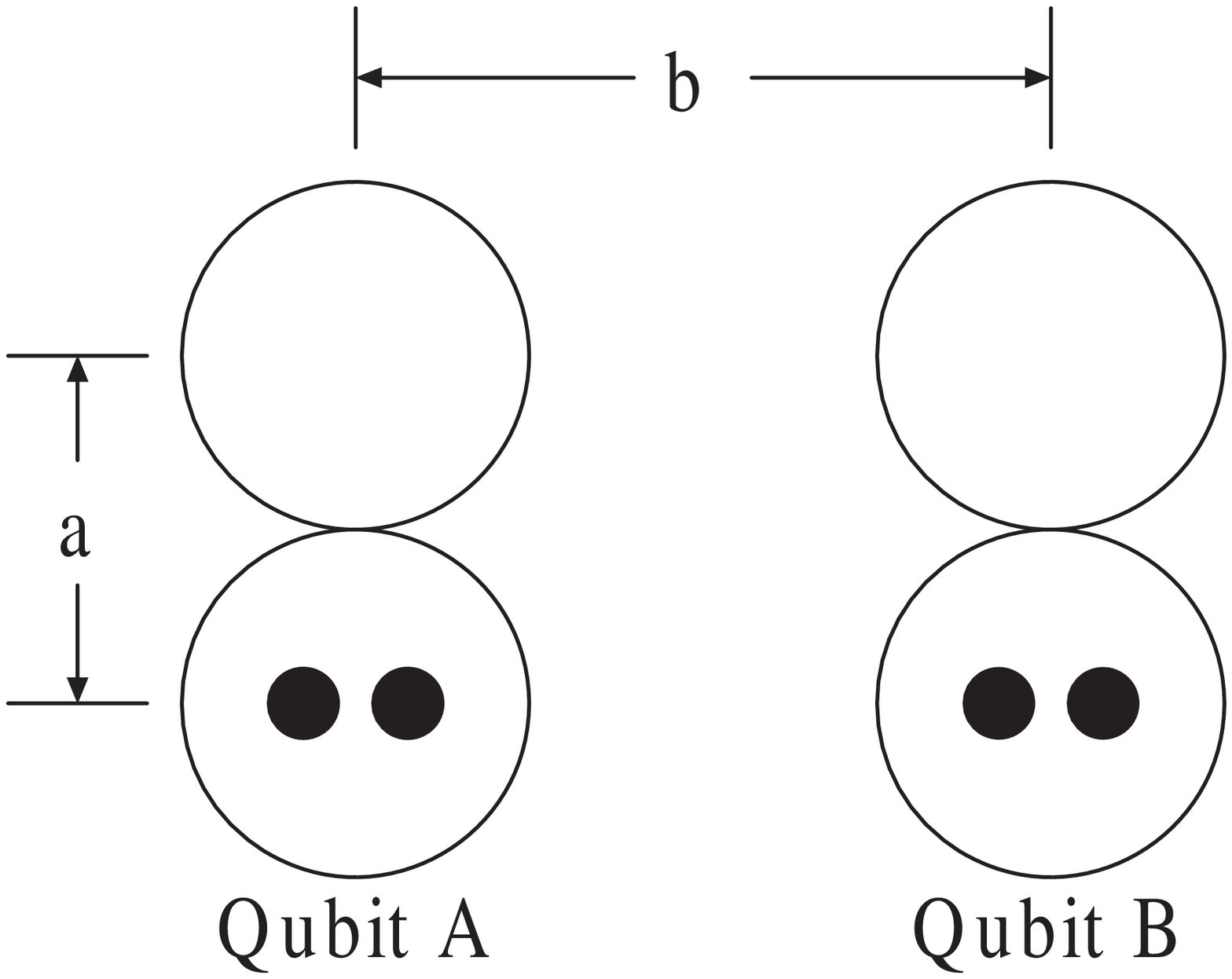}} %
\subfigure []
{\label{fig:2:d}\includegraphics[width=0.35\columnwidth]{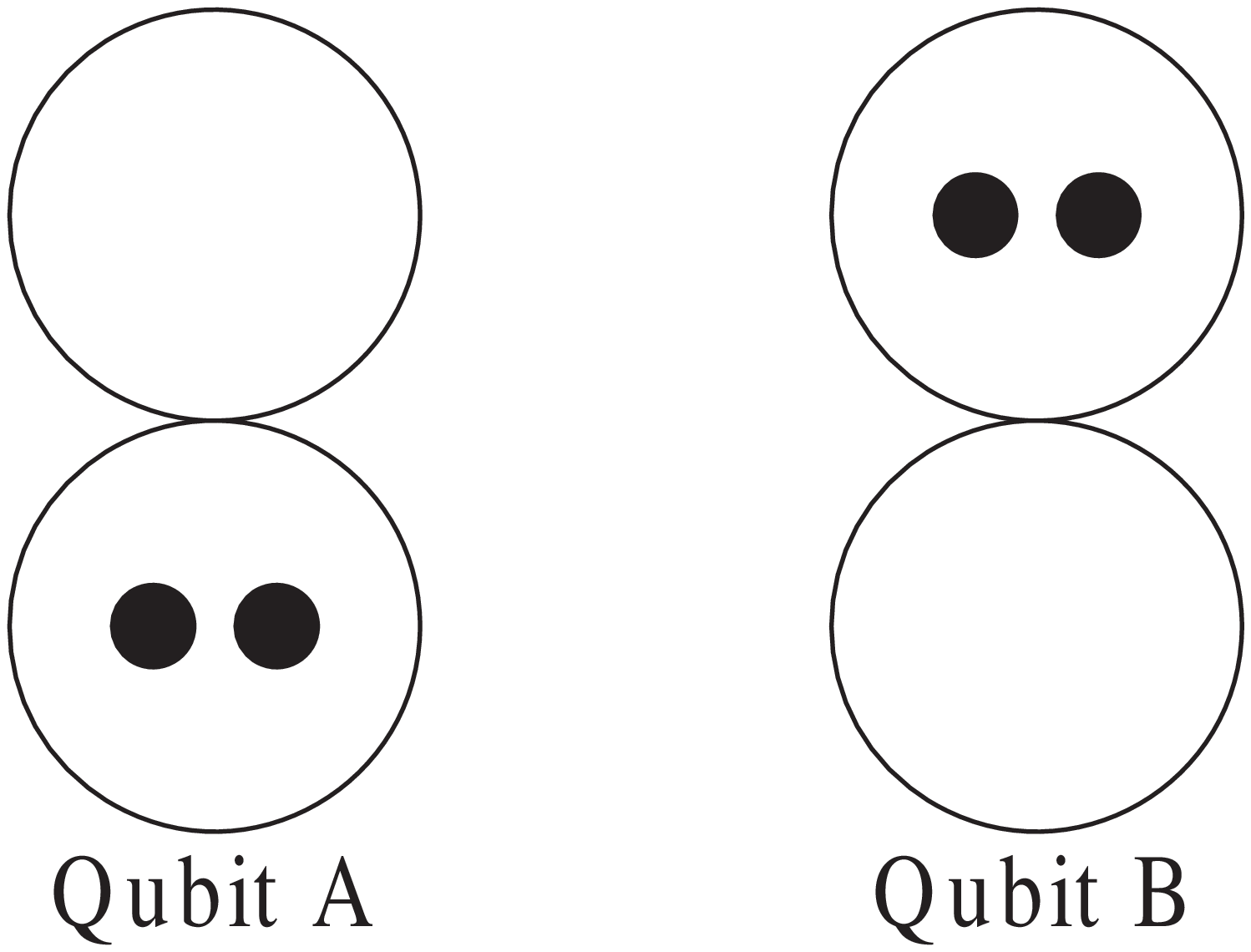}}
\caption{Relation between electron charge states and the interaction
between two QD molecules. $a$ is the distance between two QDs in one
QD molecules and $b$ is the distance between two QD molecules.
Hollow circles stand for Quantum Dots, and solid circles stand for
electrons. Every QD molecule(qubit) has two electrons. And the
interaction of two qubit depends on the electrons distribution in
the QD molecules. (a)(b) Effective interaction is zero. (c)
Effective interaction is $H_{+}$. (d) Effective interaction is
$H_{-}$. See text for details.} \label{fig:2}
\end{figure}

As the Coulomb interaction between the two electrons inside each
molecule can be excluded, we only consider the interaction between
electrons of different molecules \cite{Hanson, Stepanenko}. For any
two quantum molecules, different charge states can result in
different Coulomb interactions as shown in Fig.\ref{fig:2}. When the
bias voltage of two molecules are both set on $\Delta =0$, these two
molecules can be both in the charge state $(1,1)$ (see
Fig.\ref{fig:2:a}). In this case, the Coulomb interaction energy
between them can be directly written as:
\begin{equation}
E_{0}(a,b)=\frac{1}{4\pi \epsilon }\left( \frac{2e^{2}}{b}+\frac{2e^{2}%
}{\sqrt{a^{2}+b^{2}}}\right) .  \label{equ:E0}
\end{equation}%

Here and in the following text, $\epsilon $ is dielectric constant
of GaAs, $a$ is the distance between the two QDs in one molecule and
$b$ is the distance between the two molecules. When one of two
molecules remains in the charge state $(1,1)$ by keeping its $\Delta
=0$ and the other molecule is changed to other charge state $(0,2)$
or $(2,0)$ by sweeping its $\Delta $ to $E_{c}$
or $-E_{c}$, the interaction between them can remain in the form $%
E_{0}(a,b)$ as shown in Fig.\ref{fig:2:b}. When the two molecules
are
both in the charge state $(0,2)$ or $(2,0)$ by tuning their bias parameter $%
\Delta $, the interaction energy between them can then be written as $E_{1}(a,b)=\frac{1%
}{4\pi \epsilon }\frac{4e^{2}}{b}$ as shown in Fig.\ref{fig:2:c}. As
only the singlet state $\left\vert S\right\rangle $ can be tuned
into charge state $(0,2)$ or $(2,0)$ and the triplet state
$\left\vert T\right\rangle $ will be kept in the charge state
$(1,1)$ by sweeping $\Delta $, this interaction Hamiltonian between
two neighboring qubits can be written in the basis $\left\vert
TT\right\rangle $, $\left\vert TS\right\rangle $, $\left\vert
ST\right\rangle $ and $\left\vert SS\right\rangle $ as \cite{Guo}:
\begin{equation}
H_{+}(a,b) = \mathrm{diag}\left\{
E_{0}(a,b),E_{0}(a,b),E_{0}(a,b),E_{1}(a,b)\right\} . \label{equ:H+}
\end{equation}%
Taking the interaction energy $E_{0}(a,b)$ as zero point of the
effective interaction energy between two qubits, we can write the
interaction Hamiltonian as:
\begin{equation}
  H_{+}(a, b)=H_{0}(a, b)+H_{+}^{\prime}(a, b),
\end{equation}
here we define:
\begin{equation}
  H_{0}(a, b)=E_{0}(a, b)\times \mathrm{diag}\left\{ 1,1,1,1\right\};
\end{equation}
\begin{equation}
H_{+}^{\prime}(a,b)=\mathrm{diag}\left\{ 0,0,0,E_{+}(a,b)\right\} ,
\end{equation}%
where $E_{+}(a,b)=E_{1}(a,b)-E_{0}(a,b)=\frac{1}{4\pi \epsilon }%
\left( \frac{2e^{2}}{b}-\frac{2e^{2}}{\sqrt{a^{2}+b^{2}}}\right)$ is
positive. When one of two molecules is in the charge state $(0,2)$
and the other one is in the charge state $(2,0)$ as shown in
Fig.\ref{fig:2:d}, the interaction energy between
them can be written as $E_{2}(a,b)=\frac{1}{4\pi \epsilon }\frac{4e^{2}}{%
\sqrt{a^{2}+b^{2}}}$. Similarly, we can also get effective
interaction in the basis $\left\vert TT\right\rangle $, $\left\vert
TS\right\rangle $, $\left\vert ST\right\rangle $ and $\left\vert
SS\right\rangle $ as:
\begin{eqnarray}
H_{-}(a,b) &=& \mathrm{diag}\left\{
E_{0}(a,b),E_{0}(a,b),E_{0}(a,b),E_{2}(a,b)\right\}
\nonumber \\
&=& H_{0}(a, b)+H_{-}^{\prime}(a, b), \label{equ:H-}
\end{eqnarray}%
where
\begin{equation}
  H_{+}^{\prime}(a,b)=\mathrm{diag}\left\{ 0,0,0,E_{-}(a,b)\right\} ,
\end{equation}
and $E_{-}(a,b)=E_{2}(a,b)-E_{0}(a,b)=-E_{+}(a,b)$ is negative.

The effective interaction between two non-neighboring qubits can
also be written in the above forms, except that the parameter $b$ is
replaced by the distance between the two non-neighboring molecules.
Therefore, we can set the effective interaction between any two
quantum molecules as positive, zero and negative by properly
changing the charge state of each molecule. The property enables us
to cancel the effect of non-nearest-neighbor qubits interactions
with a multi-step preparation process for cluster state of QD
molecules.

Firstly, all qubits are initialized in the states $(\left\vert
S\right\rangle +\left\vert T\right\rangle )/\sqrt{2}$. In this case,
all molecules are in the $(1,1)$ charge state and the effective
interaction between any two qubits is closed. Sweeping $\Delta $ of
all molecules by tuning gate-bias voltages of each molecule or
applying a globe external electrical field, the qubits in the
$\left\vert S\right\rangle $ state can be changed into $(0,2)$
charge state. With this operation, Ising-model interactions as
Eq.\ref{equ:H+} are switched on between all quantum molecules. As in
our previous scheme \cite{Guo}, if there are only interactions
between nearest-neighbor quantum molecules, we can prepare cluster
state simply with this operation. However, there are small
interactions between non-nearest-neighbor QD molecules. For example,
there is interaction between qubit $i$ and qubit $i+k$ in the
following form:
\begin{eqnarray}
H_{+}(a,kb) &=& E_{0}(a,\left\vert k\right\vert b)\times
\mathrm{diag}\left\{ 1,1,1,1\right\}+ \nonumber
\\ && \mathrm{diag}\left\{ 0,0,0,E_{+}(a,\left\vert k\right\vert
b)\right\} \nonumber
\\ &=& H_{0}(a, kb)+H_{+}^{\prime}(a, kb).
\end{eqnarray}%
Here $kb$ represents the distance between molecule $i$ and $i+k$;
$k=\pm 1, \pm 2, \cdots$; when $k<0$($k>0$), $i+k$ correspond to the
qubits of left(right) side neighbor. We describe the effect of
non-nearest-neighbor qubits by the ratio of all the
non-nearest-neighbor interactions to the nearest-neighbor
interaction (based on the following discussion, the effect of
$H_{0}(a, kb)$ can be ignored):
\begin{equation}
R(a,b)=\sum_{k=2}^{+\infty }E_{+}(a,kb)/E_{+}(a,b).
\end{equation}%
For a typical value of $a$ and $b=10a$, the ratio $R(a,10a)\approx 20\%$.
Therefore, we need to include interactions between non-nearest-neighbor
qubits to generate a cluster state with higher fidelity.

Including the non-nearest-neighbor interactions, the total interaction
Hamiltonian can be written as:
\begin{equation}
H=\sum_{i=-\infty }^{+\infty
}\left(H_{+}^{i,i+1}(a,b)+\sum_{k=2}^{+\infty
}H_{+}^{i,i+k}(a,kb)\right). \label{equ:H-all}
\end{equation}%
The superscripts in Eq.\ref{equ:H-all} indicate the two qubits which
the interactions acting on. It is noted that the interaction
Hamiltonians between any two molecules are Ising-model and mutually
commute, so the order of application of the time evolution
operations does not matter. We can then describe the system time
evolution as follows:
\begin{eqnarray}
U(t) &=& exp\left(\frac{it}{\hbar}\sum_{i=-\infty }^{+\infty
}\left(H_{+}^{i,i+1}(a,b)+\sum_{k=2}^{+\infty
}H_{+}^{i,i+k}(a,kb)\right)\right)  \nonumber \\
&=&\prod_{i=-\infty ,k=1}^{+\infty }U^{i,i+k}(t)  \nonumber \\
&=&\prod_{i=-\infty ,k=2}^{+\infty }U^{i,i+k}\otimes \prod_{i=-\infty
}^{+\infty }U^{i,i+1}
\end{eqnarray}%

Here
\begin{eqnarray}
U^{i,i+k}(t)&=& exp\left(\frac{itH_{+}(a, kb)}{\hbar}\right) \nonumber \\
&=& e^{i\phi(k, t)}\mathrm{diag}\left\{1, 1, 1, exp\left( \frac{iE_{+}(a,\left\vert k\right\vert b)t}{%
\hbar }\right) \right\},
\end{eqnarray}%
where
\begin{equation}
  e^{i\phi(k, t)}=exp\left( \frac{iE_{0}(a,\left\vert k\right\vert b)t}{\hbar
  }\right),
\end{equation}
is a constant phase, which only affect the trivial global phase of
final state. That means we can ignore the $H_{0}(a, kb)$ part in
$H_{\pm }(a, kb)$. For briefness, we use $H_{\pm }(a, kb)=H_{\pm
}^{\prime}(a, kb)$

If there are only interactions between nearest-neighbor quantum
molecules, the evolution operator includes simply the term
$\prod_{i=-\infty }^{+\infty }U^{i,i+1}$, with which the system can
evolve into cluster state for a proper time $t=t_{0}$ \cite{Guo}. In
order to eliminate the effect of
non-nearest-neighbor qubit interaction, we need to cancel the operation $%
\prod_{i=-\infty ,k=2}^{+\infty }U^{i,i+k}$ (The constant  phases
caused by each $H_{0}(a, kb)$ can be added together as a global
phase, which has no effect to the final state fidelity). As shown
above, we can get negative Ising-model interaction $H_{-}(a,kb)=-$
$H_{+}(a,kb)$ by changing the molecule charge state. If the system
successively evolves under $H_{+}(a,kb)$ and $H_{-}(a,kb)$ for a
same time interval, the system will return to its origin states.
Thus we can split the cluster state preparation into several steps
to eliminate the effect of non-nearest-neighbor qubits: after the
first step operation $U(t)$, we can properly switch on negative
interaction $H_{-}(a,kb)$ to cancel the effect of the operator
$\prod_{i=-\infty ,k=2}^{+\infty }U^{i,i+k}$. The number of the
steps depends on how many non-nearest-neighbor qubit interactions we
want to cancel.

\begin{figure}[h]
\subfigure []
{\label{fig:3:a}\includegraphics[width=0.90\columnwidth]{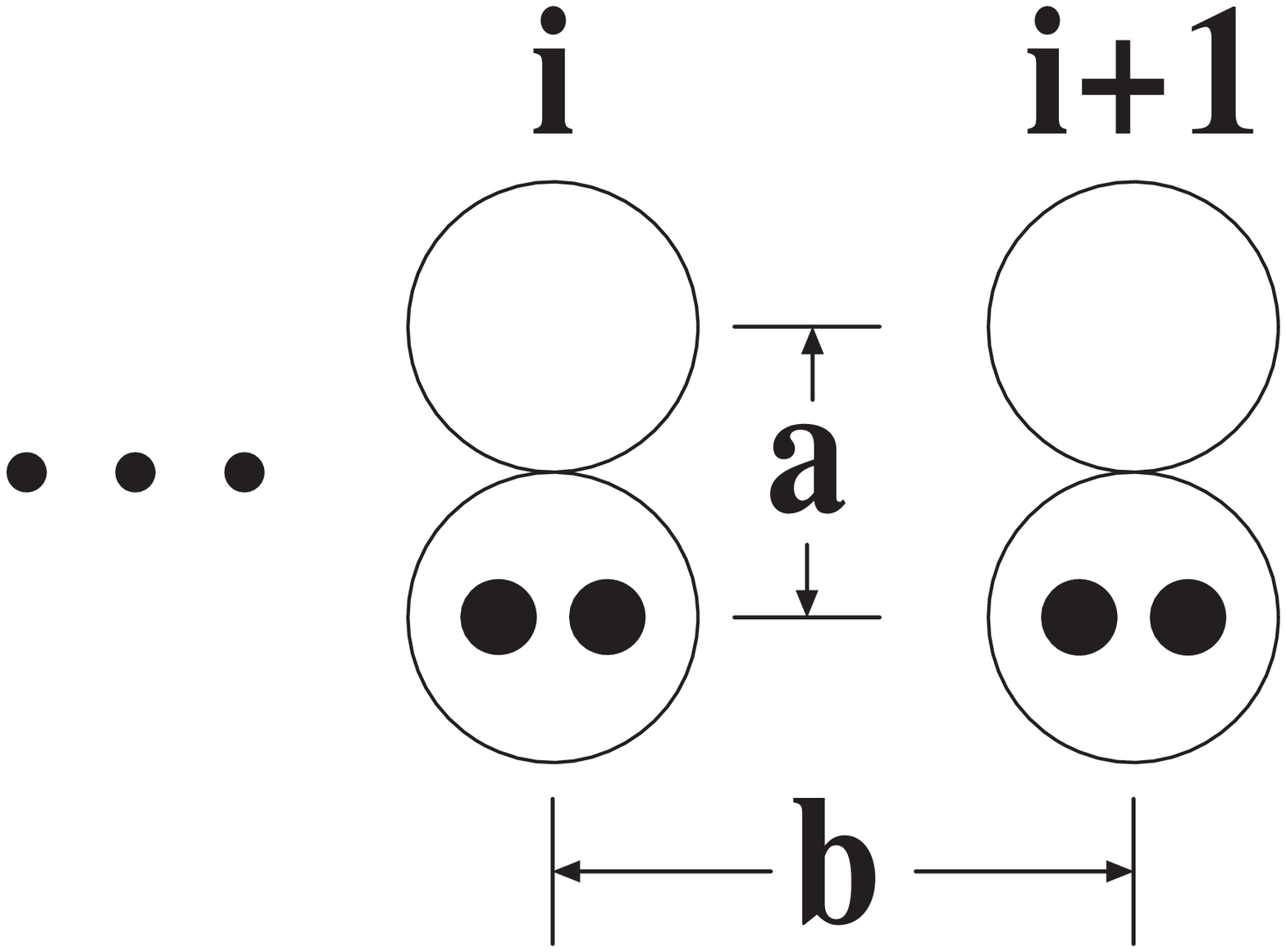}} %
\subfigure []
{\label{fig:3:b}\includegraphics[width=0.90\columnwidth]{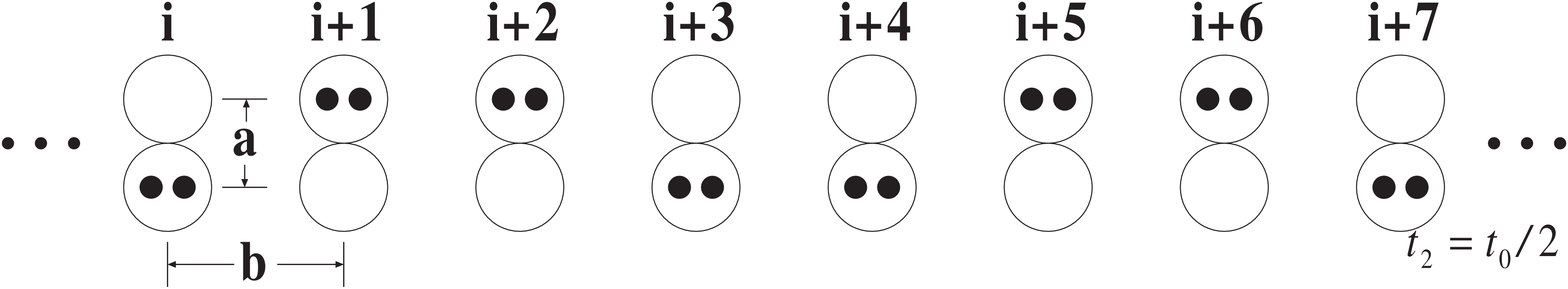}} %
\subfigure []
{\label{fig:3:c}\includegraphics[width=0.90\columnwidth]{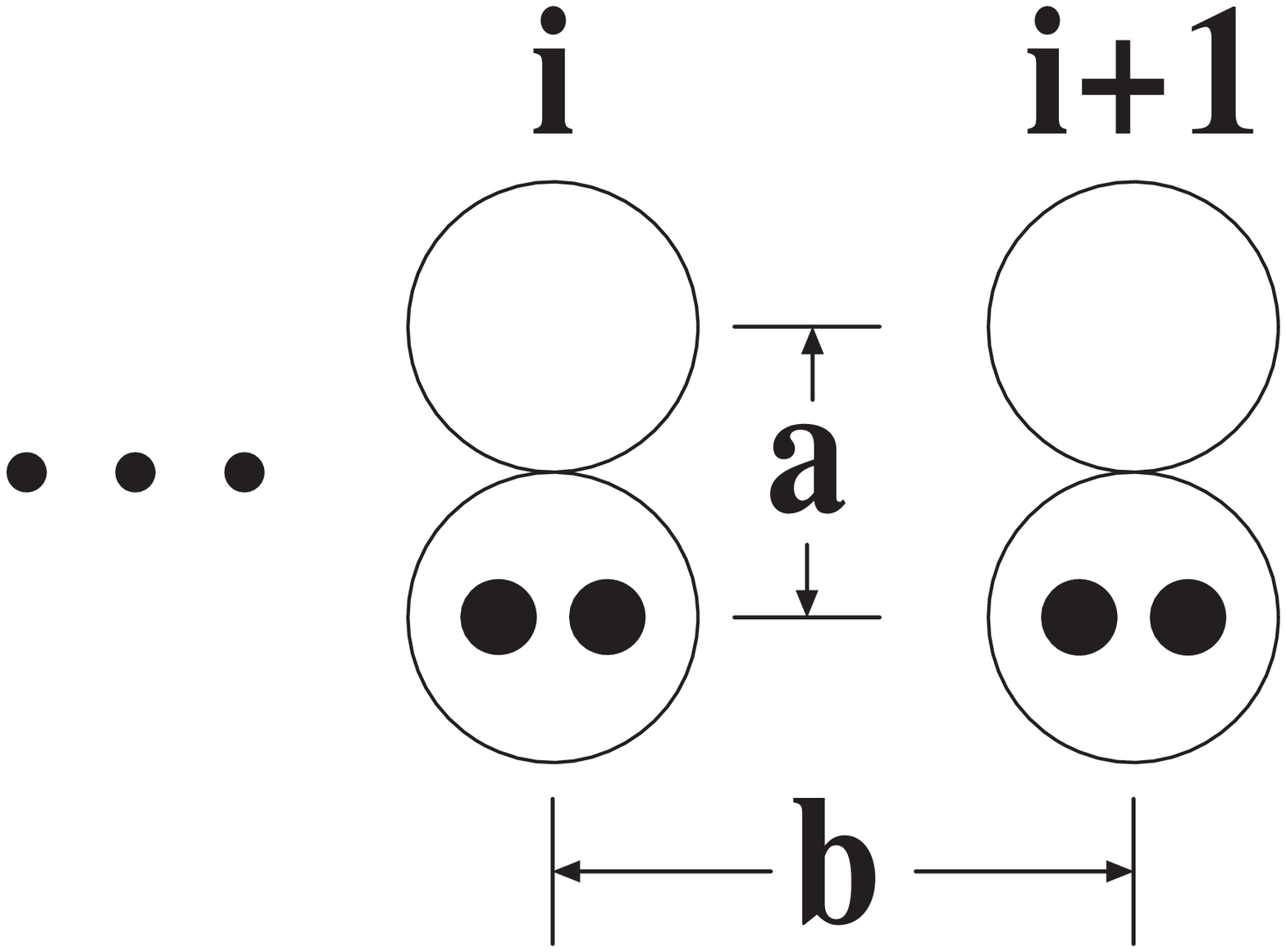}}
\caption{Scheme of eliminating the interactions between the qubits
and their next-nearest-neighbor qubits. The time shown in the right
bottom of each subfigure means the time interval of this kind of
interaction.} \label{fig:3}
\end{figure}

For example, to eliminate the effects of the next-nearest-neighbor
qubits, we can use a three-step process as given in Fig.\ref{fig:3}
and Table.\ref{table:next-n-n} to generate cluster state. Firstly,
after initialization, we tune bias parameter $\Delta $ of each
molecule to move qubits $i+4n$, $i+4n+1$ (the number $n=0,\pm 1,\pm
2,\cdots $) into the charge state $(0,2)$ and qubits $i+4n+2$,
$i+4n+3$ into the charge state $(2,0)$ for a time interval of
$t_{1}=t_{0}/2$ (Fig.\ref{fig:3:a}). Here, $ t_{0}$ is the time we
need to generate cluster state without considering the interactions
between non-nearest-neighbor qubits. Then we move the qubits $i+4n$,
$i+4n+3$ into the charge state $(0,2)$ and qubits $ i+4n+1$,
$i+4n+2$ into $(2,0)$ for a time interval of $t_{2}=t_{0}/2$
(Fig.\ref{fig:3:b}). Finally, we can move all molecules into the
charge state $(0,2)$ for a time interval of $t_{3}=t_{0}$
(Fig.\ref{fig:3:c}). Through these three steps causing a time of
$2t_{0}$, we can prepare a cluster state where the effect of all the
$(4n+2)$-nearest-neighbor qubits ($n=0,\pm 1,\pm 2,\cdots $ ),
including next-nearest-neighbor qubits ($n=0,-1$), are eliminated.
Here $(4n+2)$-nearest-neighbor means the qubit separated $4n+2$
sites from qubits $i$. By indicating the effective interaction
energies of each step in Table.\ref{table:next-n-n}, we can
understand this preparation process more clearly. Although the
interactions between qubit $i$ and qubit $i+4n$ are doubled, the
ratio of the residual non-nearest-neighbor interaction to the
nearest-neighbor interaction is reduced to the value $R^{\prime }$
through this three-step process:
\begin{eqnarray}
R^{\prime } &=&\frac{\sum_{n=1}^{+\infty }(E_{+}(a,(4n\pm
1)b)+2E_{+}(a,4nb)}{E_{+}(a,b)}  \nonumber \\
&\approx &9\%,\ (b=10a).
\end{eqnarray}

To eliminate more non-nearest-neighbor interactions, we can use a
more-step process as shown in Table.\ref{table:non-n-n}. To cancel
the effect of $(mn+2)$-nearest-neighbor to
$(mn+m-2)$-nearest-neighbor qubits ($m$ is an integer lager than
$3$, and $n=0,\pm 1,\pm 2,\cdots $), firstly, we move the qubits
$i+mn$, $i+mn+1$ into $(0,2)$ state and the others into $(2,0)$
state for a time interval of $t_{1}=t_{0}/4$ ($t_{0}$ is the same as
before). Secondly, we move the qubits $i+mn-1$, $i+mn$ into $(0,2)$
state and the others into $(2,0)$ state for a time interval of
$t_{2}=t_{0}/4$. Thirdly, we move the qubits $i+mn-2$, $i+mn-1$ into
$(0,2)$ state and the others into $(2,0)$ state for a time interval
of $t_{3}=t_{0}/4$. Similarly shift the charge state sequence one
qubit to the left for a time interval of $t_{0}/4$ as above $m$
times. After that, we move all the molecules into the charge state
$(0,2)$ for a time interval of $t_{m+1}=(8-m)t_{0}/4$. All the
effective interactions between qubit $i$ and qubit $i+k$ in this
process are shown in Table.\ref{table:non-n-n}. It is demonstrated
that after the above $m+1$ steps, only the interactions between
qubit $i$ and $i+nm\pm 1$, $i+nm$ are left (the interactions between
qubit $i$ and $i+nm$ are doubled). Thus we can prepare a cluster
state through this process, where the effect of $(mn+2)$
-nearest-neighbor to $(mn+m-2)$-nearest-neighbor qubits is
eliminated. To satisfy $t_{m+1}=(8-m)t_{0}/4\geq 0$ (the time
interval of the last step should be positive), $m$ must be no larger
than $8$. The ratio of the residual non-nearest-neighbor
interactions to the nearest-neighbor interaction is reduced to the
value $R^{\prime \prime }$:
\begin{eqnarray}
R^{\prime \prime } &=&\frac{\sum_{n=1}^{+\infty }(E_{+}(a,(mn\pm
1)b)+2E_{+}(a,mnb))}{E_{+}(a,b)}  \nonumber \\
&\approx &1\%,\ (b=10a,m=8).
\end{eqnarray}%
The total time we need to prepare cluster state in this multi-step
process is only doubled. It is noted that this above process only
can be applicable for the case $4\leq m\leq 8$. When $m=3$, it
corresponds to the case of eliminating the effect of
next-nearest-neighbor qubits. There are still some
non-nearest-neighbor interactions left in this cluster state
generation process. However, as we can turn on or off the
interaction $H_{+}(a,kb)$ or $H_{-}(a,kb)$ between any two qubits,
the remaining non-nearest-interactions can also be eliminate by a
more-step process, where the negative interactions $H_{-}(a,kb)$ are
properly switched on and off to cancel the effect of
non-nearest-neighbor interactions $H_{+}(a,kb)$. It is noted that
the present idea can only be applicable when the interaction
Hamiltonians between qubits mutually commute.

\begin{figure}[h]
\includegraphics[width=1\columnwidth]{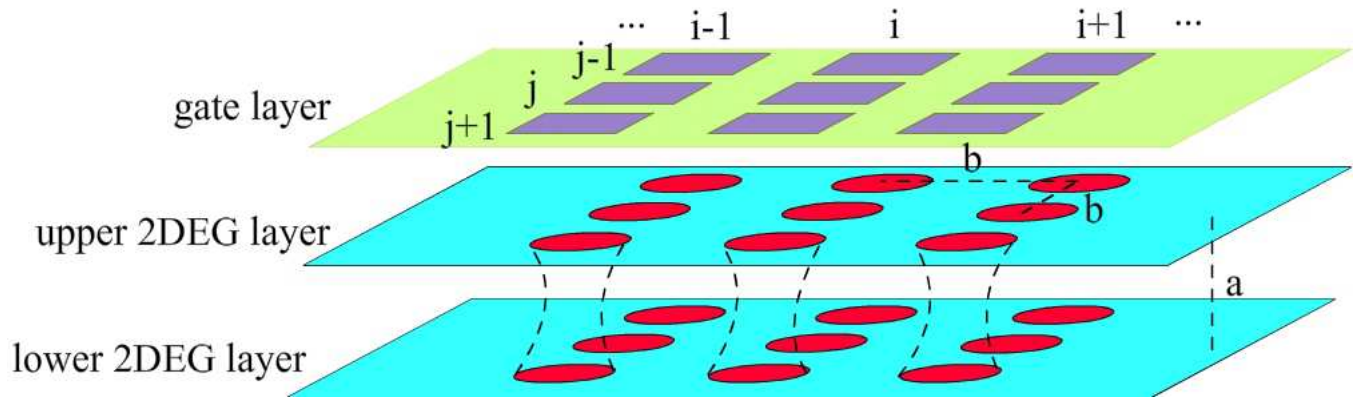}
\caption{Drawing of two-dimensional QD molecule array. Each QD
molecule is fabricated across the two layers of the system. The
squares in the gate layer stand for the control gates, and the
circles in the two-dimensional electron gas (2DEG) layers stand for
quantum dots. When the gates are tuned properly, the corresponding
qubit electron charge states can be modified.}
\label{fig:qubit-array}
\end{figure}

In the above section, only one-dimensional qubit chain is
considered. We can expand the above idea to a two-dimensional qubit
array, where the two dots in each molecule can be fabricated in
different layers of a bilayer system as shown in
Fig.\ref{fig:qubit-array} \cite{macdonald,dassarma,yamamoto,zhang}.
Each molecule can be transferred among $(0,2)$, $(1,1)$ and $(2,0)$
charge states by tuning the voltage of top gates and back gates. For
example, we can eliminate the effect of the next-nearest-neighbor
(diagonal) qubits interactions in a three-step cluster state
preparation process. Firstly, we move qubits $(i+2m,j+2n)$ to
$(0,2)$ state, qubits $(i+2m+1,j+2n+1)$ to $(2,0)$ state and the
other qubits to $(1,1)$ state for a time interval of $t_{0}$ (Here
$m,n=0,\pm 1,\pm 2,\cdots $). Secondly, we move qubits
$(i+2m,j+2n+1)$ to $(0,2)$ state, qubits $(i+2m+1,j+2n)$ to $(2,0)$
state and the others to $ (1,1)$ state for a time interval of
$t_{0}$. Finally, we add a negative voltage on the top gates to move
all the initialized qubits to $(0,2)$ charge state for a time
interval of $t_{0}$. With these three steps causing a total time of
$3t_{0}$, we can prepare a two-dimensional cluster state while the
next-nearest-neighbor (diagonal) qubits interactions are eliminated.
Similarly, with a more-step process, other non-nearest-neighbor
interactions can also be eliminated by properly switching on the
positive or negative interactions between each two qubits.

In real experiment, there are unavoidable randomness in the the
distances between quantum dots (the value of $a$ and $b$). This
randomness will greatly affect the fidelity of final state,
especially when the interactions between non-neighboring qubits are
included. As in the present protocol, the interactions of
non-neighboring qubits are canceled by switched to negative and
positive value. The randomness in the distances between
non-neighboring qubits is also eliminated.

For a wide class of so-called ``toy models'' in quantum computation,
most researches focus on the idea rather than its quantitative
development in real systems. Nevertheless, the importance and
efficiency of those simplified models are still very high, since
they usually provide physical ground for more involved theories.
Moreover, the lack of pronounced results in up-to-date experimental
efforts in realization of even prototypical quantum registers,
composed of tens of qubits, forces us to consider and to analyze all
proposed frameworks of quantum computations. Our results may
stimulate further theoretical investigations of perspectives of
one-way quantum computations for which the existence of reliable
mechanism of cluster state preparation is crucial. And it be
interesting and useful for researchers dealing with this area of
investigations.

\bigskip

In conclusion, we propose a multi-step cluster state generation
protocol for double-dot molecules to eliminate the effect of the
interactions between non-nearest-neighbor qubits. As the interaction
Hamiltonians between qubits are Ising-model and mutually commute,
the order of application of the time evolution operations does not
matter. We can get positive and negative effective interactions
between qubits to cancel the effect of non-nearest-neighbor qubits
by properly tuning the electron charge states of each QD molecule.
The total time for the present multi-step cluster state preparation
scheme is only doubled for one-dimensional qubit chain and tripled
for two-dimensional qubit array compared with the time of previous
protocol neglecting the non-nearest-neighbor interactions.

This work was funded by National Fundamental Research Program, the
innovation funds from Chinese Academy of Sciences and National Natural
Science Foundation of China (Grant No.60121503 and No.10604052).

\newpage
\begin{table}[h]
\begin{center}
\begin{tabular}{|c||c|c|c|c|c|c|c|}
\hline k & 1 & 2 & 3 & 4 & 5 & 6 & 7 \\ \hline\hline (a) & $+1/2$ &
$-1/2$ & $-1/2$ & $+1/2$ & $+1/2$ & $-1/2$ & $-1/2$ \\ \hline (b) &
$-1/2$ & $-1/2$ & $+1/2$ & $+1/2$ & $-1/2$ & $-1/2$ & $+1/2$ \\
\hline (c) & $+1$ & $+1$ & $+1$ & $+1$ & $+1$ & $+1$ & $+1$ \\
\hline\hline Total & $+1$ & $0$ & $+1$ & $+2$ & $+1$ & $0$ & $+1$ \\
\hline
\end{tabular}%
\end{center}
\caption{Interaction energies between qubit $j$ and qubit $j+k$ in
the process of eliminating the next-nearest-neighbor interactions in
preparation of cluster state(qubit $j$ is any qubit in the molecule
chain). The rows
begin with (a), (b) and (c) are corresponding to the Fig.\protect\ref%
{fig:3:a}, Fig.\protect\ref{fig:3:b} and Fig.\protect\ref{fig:3:c} in Fig.%
\protect\ref{fig:3}, respectively. And the numbers in last four
lines indicate the effective intense of the interaction between
qubit $j$ and qubit $j+k$(Units: $E_{+}(a,kb)$). }
\label{table:next-n-n}
\end{table}

\begin{table}[h]
\begin{center}
\begin{tabular}{|c||c|c|c|c|c|c|c|c|c|}
\hline k & 1 & 2 & 3 & $\cdots$ & m-2 & m-1 & m & m+1 & m+2 \\

\hline\hline (1) & + & - & - & $\cdots$ & - & - & + & + & - \\

\hline (2) & - & - & - & $\cdots$ & - & + & + & - & - \\

\hline (3) & + & + & + & $\cdots$ & + & + & - & - & - \\

\hline (4) & + & + & + & $\cdots$ & - & + & + & + & + \\

\hline $\cdots$ & \multicolumn{9}{|c|}{$\cdots$} \\

\hline (m-2) & + & + & - & $\cdots$ & + & + & + & + & + \\

\hline (m-1) & + & - & - & $\cdots$ & + & + & + & + & - \\

\hline (m) & - & - & + & $\cdots$ & + & + & + & - & - \\

\hline (1)$\sim$(m) & m-4 & m-8 & m-8 & $\cdots$ & m-8 & m-4 & m & m-4 & m-8 \\

\hline (m+1) & 8-m & 8-m & 8-m & $\cdots$ & 8-m & 8-m & 8-m & 8-m &8-m \\

\hline Total & 4 & 0 & 0 & $\cdots$ & 0 & 4 & 8 & 4 & 0 \\

\hline
\end{tabular}%
\end{center}
\caption{Interaction energies between qubit $j$ and qubit $j+k$ in
the process of eliminating the non-nearest-neighbor interactions in
preparation of cluster state (qubit $j$ is any qubit in the molecule
chain). The signs and numbers indicate the effective intense of the
interaction between qubit $j$ and qubit $j+k$(Units: $E_{+}(a,
kb)/4$). } \label{table:non-n-n}
\end{table}

\newpage

\end{document}